\documentclass[12pt]{article}
\usepackage{epsfig}
\usepackage{ifthen}
\usepackage{cite}

\begin{document}

\makeatletter \renewcommand\@biblabel[1]{(#1)}

\def\@biblabel#1{#1.}
\def\@lbibitem[#1]#2{\item[\@biblabel{#1}]\if@filesw
{\def\protect##1{\string ##1\space}\immediate
\write\@auxout{\string\bibcite{#2}{#1}}}\fi\ignorespaces}

\makeatletter \renewcommand\@cite[1]{#1} 
\def\@cite#1#2{$^{({#1\if@tempswa , #2\fi})}$}

\markright{\it Foundations of Physics\/ \bf 28 \rm 959--970 (1998)}

\vbox to 3cm{\vfill}

\voffset=-40pt
\hoffset=-50pt
\textwidth=180mm
\textheight=200mm
\begin{center}
{\LARGE\bf Realistic Interaction-Free Detection of} 
\end{center}
\smallskip
\begin{center}
{\LARGE\bf Objects in a Resonator}
\end{center}

\bigskip
\bigskip\bigskip
\parindent=2cm\hangindent=4cm\baselineskip=15pt

\bf Harry Paul\footnote{AG Nichtklassische Strahlung,  
Humboldt University of Berlin, D-12484 Berlin, Germany; 
E-mail: paul@photon.fta-berlin.de} and Mladen 
Pavi\v ci\'c\footnote{AG Nichtklassische Strahlung,  
Humboldt University of Berlin, D-12484 Berlin, Germany; 
E-mail: pavicic@w421zrz.physik.tu-berlin.de; 
Atominstitut der \"Osterreichischen Universit\"aten, 
Sch\"uttelstra\ss e 115, A--1020 Wien, Austria; 
E-mail: pavicic@ati.ac.at; Department of Mathematics, University 
of Zagreb, GF, Ka\v ci\'ceva 26, HR--41001 Zagreb, Croatia; 
E-mail: mpavicic@faust.irb.hr}
\rm

\medskip
\vrule height .3ex width 11.70493cm depth -.05ex

\parindent=2cm\hangindent=2cm\baselineskip=15pt
\medskip
We propose a realistic device for detecting objects almost 
without transferring a single quantum of energy to them. The 
device can work with an efficiency close to 100\%\ and relies 
on two detectors counting both presence and absence of the objects. 
Its possible usage in performing fundamental experiments as well 
as possible applications are discussed.   

\medskip

\vrule height .3ex width 11.71043cm depth -.05ex

\medskip



\smallskip

\vrule height .3ex width 11.71043cm depth -.05ex

\vfill\eject

\markright{H. Paul and M. Pavi\v ci\'c, Realistic Interaction-Free 
Detection\dots}

\parindent=20pt\hangindent=0pt
\baselineskip=17pt
\section{INTRODUCTION}
\label{sec:intro}
Quantum interference of individual systems has recently 
been found capable of detecting objects without transferring 
energy to them. The effect has been named \it interaction-free 
detection\/\rm\footnote{\parindent=0pt\fontsize{10pt}{11pt}
\selectfont\rm Niels Bohr would most likely argue against the 
name in the following way: ``It is true that in the measurements 
under consideration any direct mechanical interaction of the 
system and the measuring agencies is excluded, but \dots  the 
procedure of measurements has an essential influence on the 
conditions on which the very definition of the physical 
quantities in question rests\dots [T]hese conditions must 
be considered as an inherent element of any phenomenon to 
which the term ``[interaction]'' can be unambiguously 
applied.'' \protect\cite{bohr} However, the name has 
been rather unanimously accepted in the quantum parlance 
and it is likely to stay there.} and was based on the void 
detections which destroy path indistinguishability. 
In 1986 Pavi\v ci\'c\cite{my-PhD} formulated this  
in the following way. ``Consider a photon experiment shown in 
Fig.~\ref{fig:my-PhD-exp} which results in an interference in 
the region \it D\/ \rm provided we do not know whether it arrived to the 
region by path $s_1$ or by path $s_2$. As it is well-known, 
experimental facts are: If we, after a photon passed 
the beam splitter \it B\/ \rm and before it could reach the point \it 
C\/\rm, suddenly introduce a detector in the path $s_2$ in the point 
\it C\/ \rm and do \it not\/ \rm detect \it anything\/\rm, then it follows 
that the photon must have taken the path $s_1$---and, really, 
one can detect it in the region \it D\/ \rm but it does \it not\/ 
\rm produce interference there.  Quantum mechanically, if we registered  
the interference in the region \it D\/\rm, we could not find an 
experimental procedure to directly either prove or disprove that the 
photon uses both paths simultaneously. However, the fact that by detecting 
\it nothing\/ \rm in point \it C\/ \rm we destroy the interference implies 
that the photon \it somehow\/ \rm knows of the other path 
when it takes the first one.'' (Ref.~\citen{my-PhD}, pp.~31,$\>$32) 
 
Photon's ``knowledge'' about the other path one can employ 
to detect an object (at point \it C\/\rm) without transferring even a 
single quantum of energy to it. The efficiency of such an application 
with symmetrical Mach-Zehnder interferometer (shown in 
Fig.~\ref{fig:my-PhD-exp}) is ideally only 25\%\ for 
single detections and 33\%\ in the long run as shown in Elitzur 
and Vaidman's detailed formulation of the void detections in 
interference experiments in 1993.\cite{bomb} They also showed that one 
could increase the ideal efficiency to 50\%\ if an asymmetrical beam 
splitter were used. In 1995 \rm Kwiat \it et al.\/\rm\cite{z95} carried 
out Elitzur and Vaidman's proposal with an asymmetrical beam splitter 
using  photons obtained in a parametric down conversion. In this way an 
efficiency close to 50\%\ has been achieved for correlated photons. 
However, the realization was concerned only with the confirmation
of the effect and the 50\%\ efficiency referred to the detected 
photons which supported the confirmation. For, in the experiment it 
was necessary to select, with irises, a very small fraction of the 
photons originally produced in downconversion, which resulted in a net 
detection efficiency of only 2\%. The latter efficiency can be 
significantly improved\cite{I-o-c} but the downconversion can hardly be 
used for a straightforward realistic interaction-free device.       

A proposal put forward by Kwiat \it et al.\/\rm\cite{z95,zzz95} in 
1995 which aims at realistic efficiencies of \it not hitting\/ \rm 
tested objects is shown in Fig.~\ref{fig:2-cavites}. The device 
consists of two coupled resonators (cavities) separated by a highly 
reflective beam splitter and assumes inserting single photons into 
one of them. If an object were in the other cavity the probability 
of it being hit would remain comparatively low. In the absence of 
an object the photon should, after a certain number of cycles $N$,  
be in the right cavity with certainty. Inserting of a detector 
in the left cavity should verify the cases. Such an experiment 
would be very hard to carry out in realistic conditions even if the 
problem of inserting single photons and the detector were solved. 
In particular because no firing of the detector should mean the 
absence of the object and because of the high losses at the mirrors. 

In 1996 \rm Kwiat \it et al.\/\rm\cite{z96} put forward another 
proposal, shown in Fig.~\ref{fig:zeno}, which is based on a previous 
elaboration of the optical Zeno effect. A horizontally polarized 
photon enters the resonator through the switchable mirror \it SM\/ \rm 
which keeps it in for $N$ cycles. After each cycle the polarization 
rotator \it PR\/ \rm turns the initial photon polarization by an angle 
$\alpha$. When there is no object in the resonator the wave 
function recombines at the polarizing beam splitter \it PBS\/ \rm within 
each cycle so that after $N=\pi/(2\alpha)$ cycles it exits the 
resonator vertically polarized. When there is an object in the 
resonator within each cycle we have got the Malus probability 
$p=\cos^2\alpha$ of photon passing straight through the 
horizontally polarizing beam splitter. After $N$ cycles the 
photon---horizontally polarized---exits through \it SM\/ \rm with the 
probability $P=p^N$. The probability of the object being hit by 
the photon is therefore $Q=1-P$. For $\alpha=1^\circ$, $Q=3$\%. 
In this proposal, as opposed to the previous one, we do have 
different detectable outcomes for presence and absence of objects. 
Nevertheless, one has to start again from photon pairs generated in 
a parametric downconversion in order to be able to determine the 
photon's entrance time and thus fix the number of cycles, i.e., the 
moment in which one should let the photon out of the resonator through 
the switchable mirror $SM$. Moreover, the mirror losses will have a 
detrimental effect on the experiment. Actually, this effect grows with 
the number of cycles: the larger the latter is, the lower is the ideal 
theoretical value of $Q$, and the bigger are the losses.  

In 1997 we conceived a different approach using a single 
monolithic total-internal-reflection resonator (MOTIRR) coupled 
by two frustrated-total-in\-ternal-reflection (FTIR) 
prisms.\cite{P-I-97} The physical principle of the device was  
essentially the same as for the scheme in Fig.~\ref{fig:p-b} we 
are going to present in this paper with the only difference that the 
central loops were confined within a monolithic crystal. The presence 
of the object causes firing of detector \it D\rm$_r$ and the absence 
causes firing of \it D\rm$_t$. The losses in a MOTIRR are extremely 
low and go down to 0.3\%.\cite{motirr2} Thus  a realistic 
application of interaction-free measurements to the suitable small 
objects has been enabled. In this paper we are proposing a general 
purpose interaction-free device for all possible applications  
foreseen so far, calculate the losses that can be expected in 
realistic conditions, and show that the present setup evades 
limitations of the previous proposals.  

Suggested applications of the interaction-free measurements 
are numerous and range from the foundational physical experiments  
and experiments to medicine. Let us just cite some of them.   
It has been used to show that under a plausible condition 
Lorentz-invariant realistic interpretations of quantum mechanics 
are not possible.\cite{hardy} A preparation of a very well localized 
atom beam by means of a Mach-Zehnder interferometer for neutrons 
without physical interaction has been proposed\cite{hans} and 
the first interaction-free experiment with neutrons has already 
been carried out\cite{hans-e}. A possible interaction-free 
experiment in quantum dot systems has been discussed.\cite{q-dot}  
An optical device for erasing fringes of atom interference without 
disturbing either the spatial wave function or its phase has 
been proposed\cite{IF-PLA} thus strengthening the result of 
Scully \it et al.\/\rm\cite{scully} Cf.~also Ref.~\citen{sweeden}. 
Testing of Bose-Einstein condensates (which can be blown apart by 
even a single photon) has been recently seen as the most immediate 
possible application.\cite{z96} Also a preparation of a 
superposition at a macroscopic scale.\cite{z96} And in the end 
several more distant possible applications such as   
selecting particular bacteria without killing them, safe X-ray 
photography, quantum computer application, etc.\cite{vaidman}
In any case we share the feeling that ``the situation resembles 
that of the early years of the laser when scientists knew it would 
be an ideal solution to many unknown problems.''\cite{z96}

\section{EXPERIMENT}
\label{sec:experiment}

Fig.~\ref{fig:p-b} shows an outline of the proposed experiment. 
When there is no object in the device, an incoming laser beam is 
almost totally transmitted (up to 98\%) into detector \it D\rm$_t$ 
and when there is an object, an incoming laser beam is 
being (ideally) totally reflected into detector \it D\rm$_r$. The 
device consists of four prisms forming a resonator. The prisms 
are designed so that their entrance and exit faces are at right 
angles to the beam making rectangular loops and are covered with 
multilayer antireflection coating to minimize reflection losses. 
The entrance prism is coupled to the adjacent loop prism by the 
frustrated total reflection, which is an optical version of 
quantum mechanical tunnelling.\cite{motirr2} 
Depending on the dimension of the gap between the prisms one can well 
define reflectivity $R$ within the range from $10^{-5}$ to 0.99995.
The uniqueness of the reflectivity at the gaps and at the same time 
no reflectivity at the entrance and exit faces of the prisms for each 
photon is assured by choosing the orientation of the polarization of 
the incoming laser beam perpendicular to the plane of incidence. 
As a source of the incoming beam a continuous wave laser (e.g., Nd:YAG) 
should be used because of its coherence length (up to 300$\>$km) and of 
its very narrow linewidth (down to 10$\>$kHz in the visible 
range).\cite{laser}  

Let us now determine the intensity of the beam arriving at detector 
\it D\rm$_r$ when there is no object in the path. Our detailed 
calculations\cite{I} show that a rigorous description of the 
device is formally equivalent to a Fabry-Perrot-type of a 
resonator with standard mirrors up to the phase shifts at 
the FTR's which we take into account so as to include it into 
the phase which is being added by each round-trip. The portion 
of the incoming beam of amplitude $A(\omega)$ reflected at the \it 
FTR\/ \rm inner face of the incoming prism is described by the 
amplitude $B_0(\omega)=-A(\omega)\sqrt{R_1}$, where $R$ is reflectivity. 
The remaining part of the beam tunnels into the resonator and travels 
around the resonator guided by one frustrated total reflection (with 
reflectivity $\sqrt{R_2}$ at the face 
next to the right prism where a part of the beam tunnels out into 
\it D\rm$_t$) and by two proper total reflections. The losses for such 
a set-up---as opposed to standard mirror Fabry-Perrot resonators---are 
very low as calculations and recent experiments show: below  
2\%\cite{2perc} for the type presented here and even below  
0.3\%\cite{motirr2} for the set-up with a 
monolithic resonator we presented in Ref.~\citen{P-I-97}. The losses 
in the present set-up are mostly due to absorption and scatter in the 
multilayer antireflection coatings and the crystals and to a much 
smaller extent due to imperfect total reflections. After a full 
round-trip the following portion of the beam joins the directly 
reflected portion of the beam by tunnelling into the left prism:
$B_1(\omega)=A(\omega)\sqrt{1-R_1}\sqrt{R_2}
\sqrt{R_3}\sqrt{R_4}\sqrt{1-R_1}\>e^{i\psi}$, 
where $\psi=(\omega-\omega_{res})T$ is the phase added by each 
round-trip which also includes phase shifts at the gaps; 
here $\omega$ is the frequency of the incoming beam, $T$ is the 
round-trip time, $\omega_{res}$ is the resonator frequency, and 
$\sqrt{R_3}$, $\sqrt{R_4}$ are the two (realistic, and therefore not 
equal 1) total reflectivities in which we also include the 
afore mentioned absorption and scatter (which can be treated as 
trasmitivities); here we introduce $\rho=\sqrt{R_3R_4}$ as a 
measure of all the losses; $\rho=1$ corresponds to an ideal 
case with no losses.       
   
Each subsequent round-trip contributes to the geometric progression: 
\begin{eqnarray}
B(\omega)&=&A(\omega)\{-\sqrt{R_1}+(1-
R_1)\rho\sqrt{R_2}e^{i\psi}[1+\rho\sqrt{R_1R_2}e^{i\psi}+
\dots]\}\nonumber\\
&=&A(\omega)\{-\sqrt{R_1}+{(1-R_1)\rho\sqrt{R_2}e^{i\psi}\over 
1-\rho\sqrt{R_1R_2}e^{i\psi}}\}\,,
\label{eq:total} 
\end{eqnarray}
so as to yield the following probability of the beam being 
reflected into $D_r$
\begin{eqnarray}
B(\omega)B(\omega)^*=A(\omega)A(\omega)^*[1-{(1-R_1)(1-\rho^2 R_2)\over 
1-2\rho\sqrt{R_1R_2}\cos\psi + \rho^2 R_1R_2}]\,.
\label{eq:int-r} 
\end{eqnarray}

In an analogous way we obtain the probability of the beam 
being transmitted into $D_t$
\begin{eqnarray}
C(\omega)C(\omega)^*=A(\omega)A(\omega)^*{(1-R_1)(1-R_2)\over 
1-2\rho\sqrt{R_1R_2}\cos\psi + \rho^2 R_1R_2}\,.
\label{eq:int-t} 
\end{eqnarray}

Since the frequency of the input laser beam can never precisely 
match the resonance frequency we make use of a Gaussian wave 
packet $A(\omega)=A\exp[-{\cal T}^2 (\omega-\omega_{res})^2/2]$,
where ${\cal T}$ is the coherence time which obviously must be
significantly longer than the round trip time $T$. Thus we 
describe the incident wave by 
\begin{eqnarray}
E^{(+)}_i(z,t)=\int^\infty_0A(\omega)e^{i(kz-\omega
t)}d\omega\,,
\label{eq:e}
\end{eqnarray}
the reflected wave by:
\begin{eqnarray}
E^{(+)}_r(z',t)=\int^\infty_0B(\omega)e^{i(kz'-\omega
t)}d\omega\,,
\label{eq:er}
\end{eqnarray}
and the transmitted wave by:
\begin{eqnarray}
E^{(+)}_t(z',t)=\int^\infty_0C(\omega)e^{i(kz'-\omega
t)}d\omega\,,
\label{eq:et}
\end{eqnarray}

The energy of the incoming beam is the energy flow integrated
over time:
\begin{eqnarray}
I_i=\int^\infty_{-\infty}E^{(+)}_i(z,t)E^{(-)}_i(z,t)dt=
\int^\infty_0 A(\omega)A^*(\omega)d\omega\,.
\label{eq:i}
\end{eqnarray}
The energies of the reflected and transmitted beams are given 
analogously by
$I_r=\int^\infty_0 B(\omega)B^*(\omega)d\omega$ 
and $I_t=\int^\infty_0 C(\omega)C^*(\omega)d\omega$, 
respectively. 

The efficiency of the suppression of the reflection into $D_r$ 
is given by 
\begin{eqnarray}
\eta=1-{I_r\over I_i}=(1-R_1)(1-\rho^2 R_2)\,\Phi
\,,\label{eq:eta}
\end{eqnarray}
and the efficiency of the throughput into $D_t$ by: 
\begin{eqnarray}
\tau={I_t\over I_i}=(1-R_1)(1-R_2)\,\Phi
\,,\label{eq:tau}
\end{eqnarray}
where 
\begin{eqnarray}
\Phi=
{\displaystyle\int_0^\infty{\displaystyle\exp[-
{\cal T}^2(\omega-
\omega_{res})^2/2]d\omega\over\displaystyle1-
2\rho\sqrt{R_1R_2}\cos[(\omega-
\omega_{res}){\cal T}/a]+\rho^2R_1R_2}\over\displaystyle\int_0^\infty
\exp[-{\cal T}^2(\omega-\omega_{res})^2]d\omega}
\,,\label{eq:phi}
\end{eqnarray}
where $a\equiv {\cal T}/T$ is a ratio of the coherence time 
${\cal T}$ and the round-trip time $T$. The coherence 
length should always be long enough ($a>200$) to allow
sufficiently many round trips (at least 200). $\Phi$ turns out 
to be very susceptible to the small changes of $\rho$ so as 
to yield rather different outputs of $\tau$ in opposition to
$\eta$. (Cf.~Figures \ref{fig:eta} and \ref{fig:tau}.) 
 
Obviously both $\eta$ and $\tau$ should be as close to 1 as 
possible. A computer optimization shows that this can best be 
achieved by taking $R_1=R_2$. In Figures~\ref{fig:eta} and 
\ref{fig:tau} we give the values of $\eta$ and $\tau$, respectively, 
for $\rho$'s which correspond to the throughput $\tau$ of about 98\%\ 
which is considered achievable. The total reflectivities with losses 
below $10^{-6}$ are achievable so that the given values for $\rho$ are 
not the problem so far as the total reflection is considered. As for 
the throughput $\tau$ the given values for $\rho$ are also apparently 
achievable. If however the absorption of antireflection coating   
turns out to be too high one can always substitute Pellin-Broca 
prisms with entrance and exit faces at Brewster's angles 
(i.e., no reflection losses) for the present prisms with the 
multilayer antireflection coatings.\cite{I}.   

In order to carry out the experiment we have to lower the intensity 
of the beam until it is likely that only one photon would appear 
within an appropriate time window (1$\>$ms -- 1$\>\mu$s $<$ coherence 
time) what allows the intensity in the cavity to build up. The values 
for $1-\eta$ are probabilities of detector $D_r$ reacting when there 
is no object in the system. The values for $\tau$ are probabilities of 
detector $D_t$ reacting when there is no object in the system. 
For example, for $R=0.98$ and $\rho=0.9999$ one obtains 
$\eta=0.99$ and $\tau=0.98$. $\eta$ and $\tau$ in Figs.~\ref{fig:eta} 
and \ref{fig:tau} are calculated for $a=500$, i.e., for 500 round-trips 
which are multiply assured by continuous wave laser coherence length. 
Since we did take possible background counts into account by using 
the Gaussians for the calculation, we can equally 
rely on $D_r$ and on $D_t$ firing; also, we can use this fact for 
tuning the device. A response from $D_r$ means that there is an 
object in the system. In the latter case the probability of the 
$D_r$ response is ideally $R$, the probability of a photon hitting the 
object is $R(1-R)$, and the probability of photon exiting into $D_t$ 
detector is $(1-R)^2$. 

We start each testing by opening a gate for the incident beam and 
after either \it D\rm$_r$  or \it D\rm$_t$ fires, the 
testing is over. The cases when detectors fail to react either because 
of their inefficiency are not problematic because single photon detectors 
with 85\%\ efficiency are already available. Such a failure 
would result in a slightly bigger time window, so that a chance of a 
photon hitting an object would remain practically unchanged.
Thus, a possible 300$\>$km coherence length of cw lasers does not leave 
any doubt that a real experiment of detecting objects (with an efficiency 
of over 98\%) and without transferring a single quantum of energy to them 
(with the same efficiency) can be carried out successfully.

\section{CONCLUSION}
\label{sec:conclusion}

We have shown that with our resonator based on total reflections 
and frustrated total reflections, interaction-free measurements 
can be carried out with a realistically achievable efficiency of 
98\%. The proposed design makes the device not only very 
suitable for the foundational experiments reviewed in 
Sec.~\ref{sec:intro} but also a good candidate for a more 
general application, e.g., in medicine, for X-raying patients 
practically without exposing them to radiation. The latter 
application was not possible with previous setups because they 
were all based on mirrors and the losses at X-ray mirrors could 
be too high for building a realistic interaction-free device. 
Total reflections we use are however applicable to X-rays and 
often used for constructing X-ray lasers. On the other
hand our setup with two outputs is easily applicable to an 
interaction-free detection of \it gray\/ \rm objects where 
one concludes on the level of \it grayness\/ \rm by means 
of the statistics of repeated testings.    

\bigskip
\bigskip
\parindent=0pt
{\Large\bf ACKNOWLEDGMENTS}

\parindent=20pt
\bigskip

M.~P. acknowledges supports of the Alexander von Humboldt 
Foundation, Bonn, Germany, the Technical University of Vienna, 
Vienna, Austria,  and the Ministry of Science of Croatia. He would also 
like to thank Johann Summhammer, Atominstitut der 
\"Osterreichischen Universit\"aten, Vienna for many valuable 
discussions.   

\vfill\eject
\parindent=0pt
\vbox to 2cm{\vfill}
%{\Large\bf REFERENCES}

\bigskip

%\everypar={\parindent=0pt\hangindent=20pt\hangafter=1}

\renewcommand\refname{REFERENCES}

\vfill\eject

\begin{figure}
\begin{center}
{\Large\bf FIGURES}
\bigskip\bigskip\bigskip
\end{center}
\caption{Figure taken from Pavi\v ci\'c (1986).  
``By detecting \it nothing\/ \rm in the point \it C\/ \rm we destroy 
the interference [in the region \it D\/\rm].'' 
(Ref.~\protect\citen{my-PhD}, p.31)}
\label{fig:my-PhD-exp}
\bigskip
\caption{Figure according to Ref.~\protect\citen{z95}. A single 
photon inserted into the left cavity stays there when there is 
an object in the right cavity and moves to the right cavity when 
there is no object there.}
\label{fig:2-cavites}
\bigskip
\caption{Figure according to Ref.~\protect\citen{z96}. A single 
horizontally polarized photon enters the resonator through the 
switchable mirror \it SM\/ \rm and passes the polarization rotator 
\it PR\/ \rm (which turns the polarization plane by the angle 
$90^\circ/N$) and the polarizing beam splitter \it PBS\/ \rm $N$ 
times before exiting through \it SM\/ \rm horizontally polarized 
when there is an object in the path and vertically polarized when 
there is no object in the path.}
\label{fig:zeno}
\bigskip
\caption{Schematic of the proposed realistic interaction-free 
device. A single p-polarized photon tunnels (frustrated 
total reflection, \it FTR\/\rm) into the resonator. 
With a realistic efficiency exceeding 98\% 
the beam makes several hundred loops guided by two total 
reflections \it TR\/ \rm and two \it FTR\/\rm's to exit into 
\it D\rm$_t$ when there is no object in the path. When there is 
an object in the resonator, the beam is reflected into \it D\rm$_r$.}
\label{fig:p-b}
\bigskip
\caption{The efficiency of the suppression of the reflection into $D_r$ 
when there is no object in the resonator as given by 
Eq.~\protect\ref{eq:eta}. $R$ is the frustrated total reflection 
at the two coupling output prisms and $\rho$ is the measure of 
losses as defined for Eq.~\protect\ref{eq:total}.    
}
\label{fig:eta}

\bigskip
\caption{The efficiency of the throughput into $D_t$ 
when there is no object in the resonator as given by 
Eq.~\protect\ref{eq:tau}. $R$ and $\rho$ are defined as in 
Fig.~\protect\ref{fig:eta}.
}
\label{fig:tau}
\bigskip
\vbox to 7cm{\vfill}
\end{figure}

\eject 

\vfill

\begin{figure}
\includegraphics{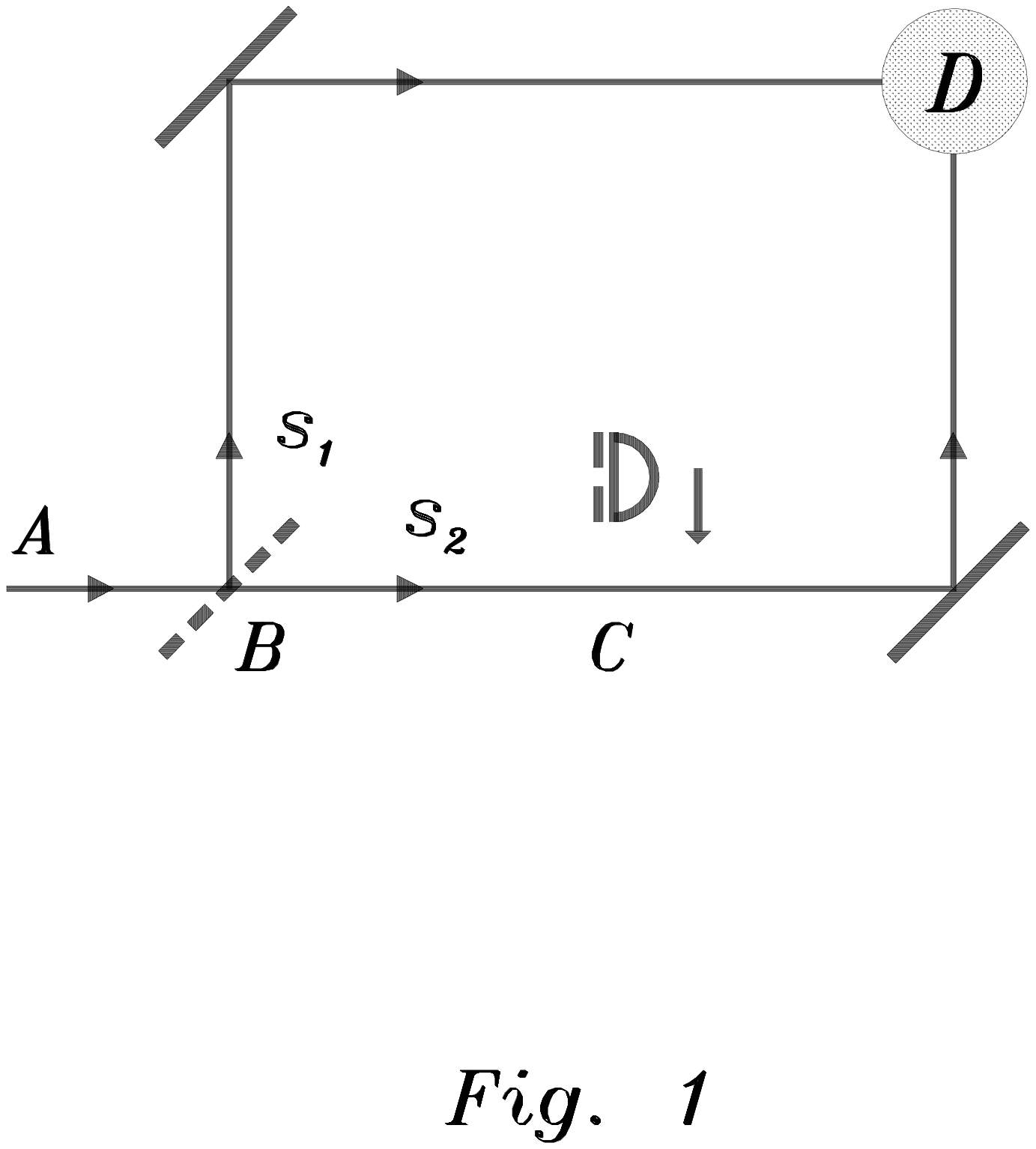}
\end{figure}

\vfill

\eject

\vfill 

\begin{figure}

\includegraphics{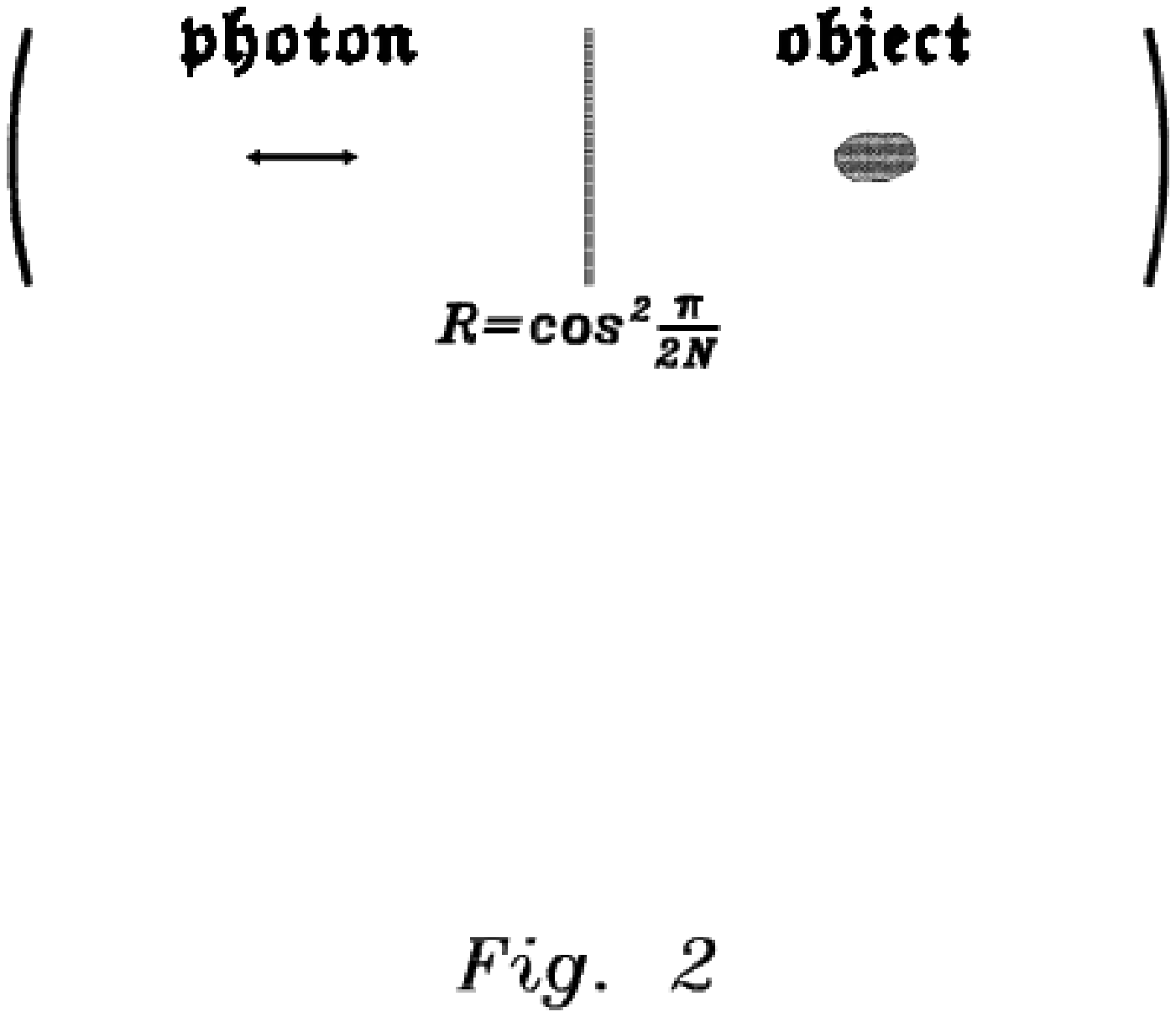}

\end{figure}

\vfill

\eject

\vfill 

\begin{figure}

\includegraphics{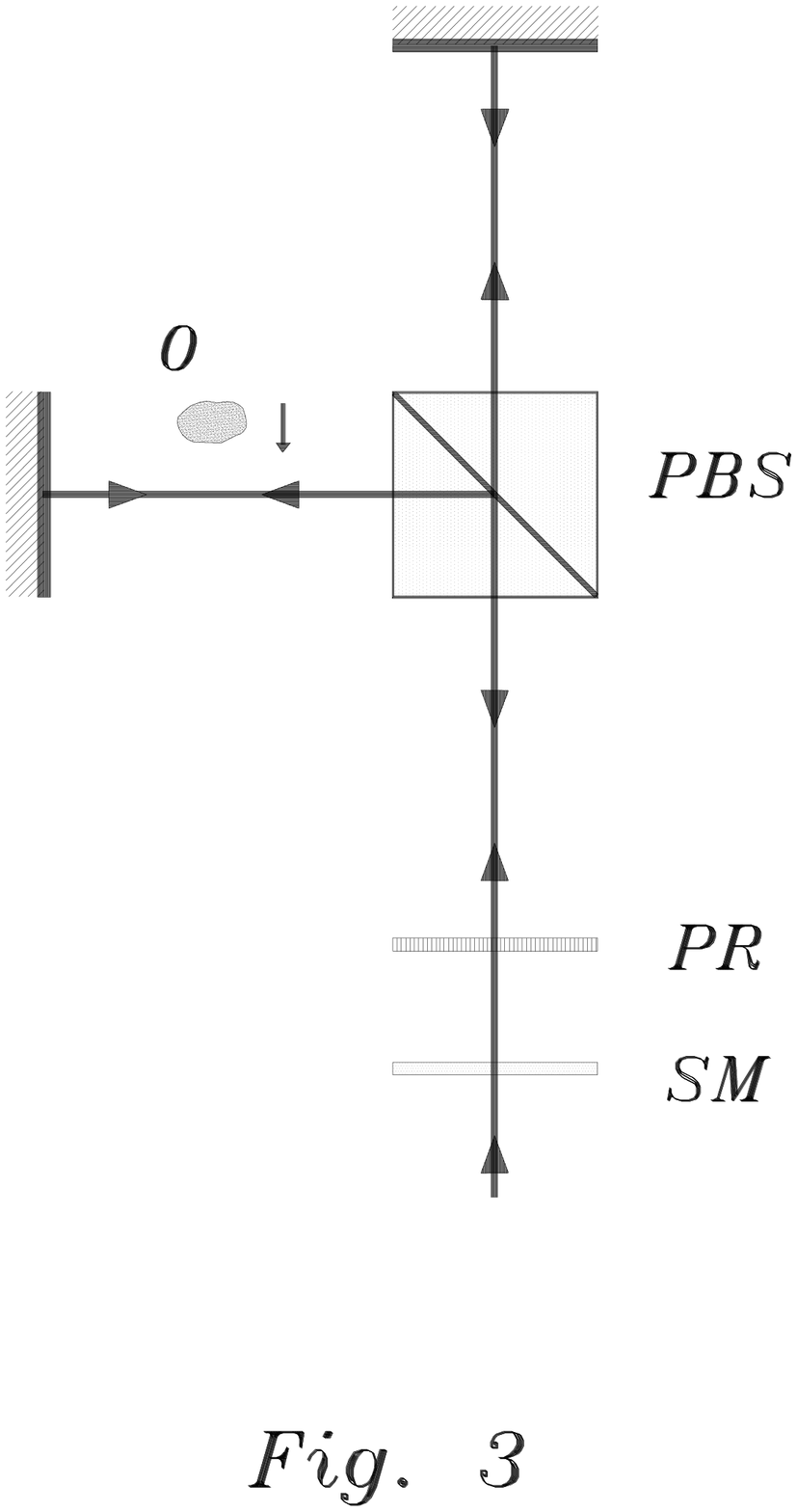}

\end{figure}

\vfill

\eject 

\vfill

\begin{figure}

\includegraphics{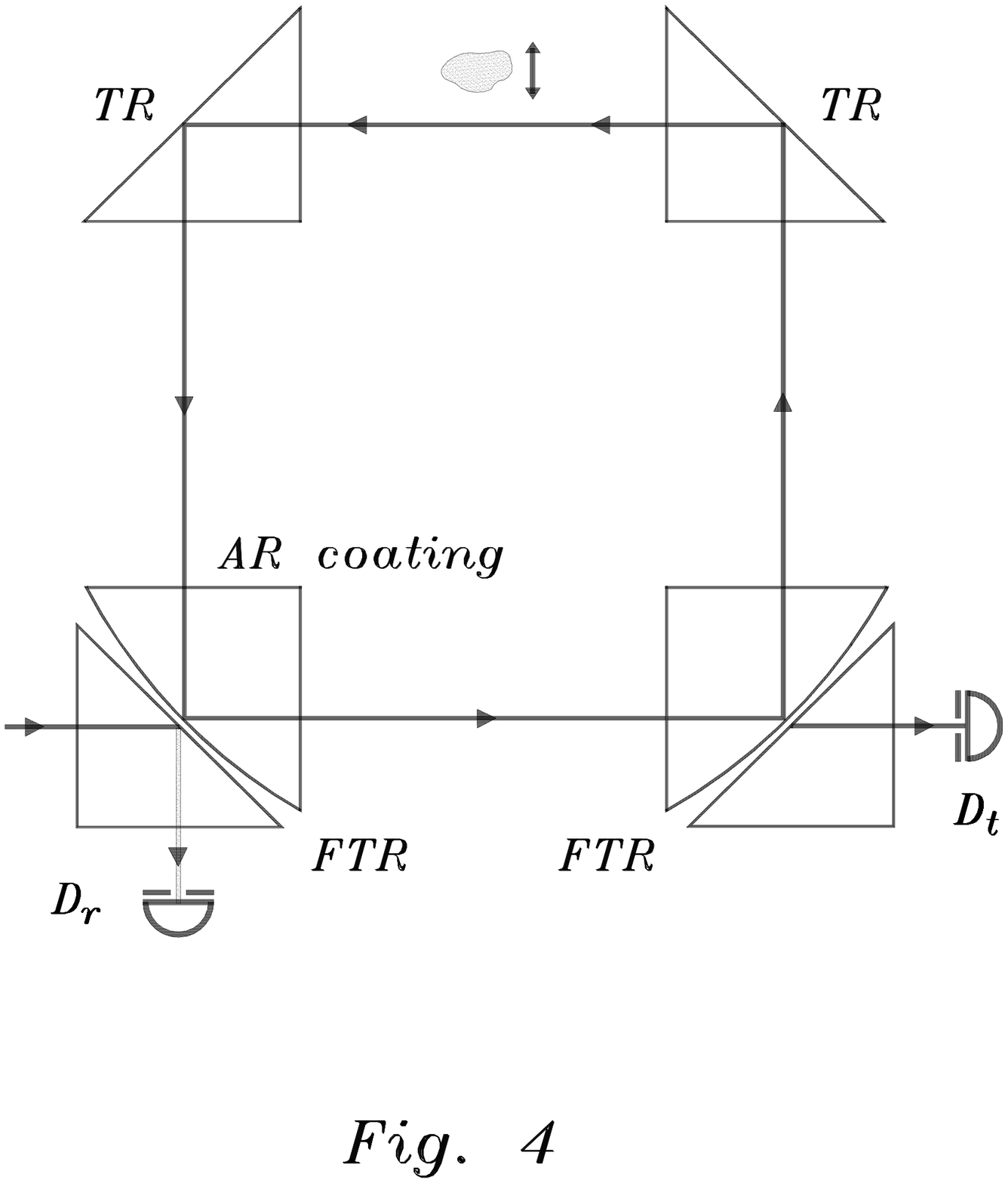}

\end{figure}

\vfill

\eject 

\vfill

\begin{figure}

\includegraphics{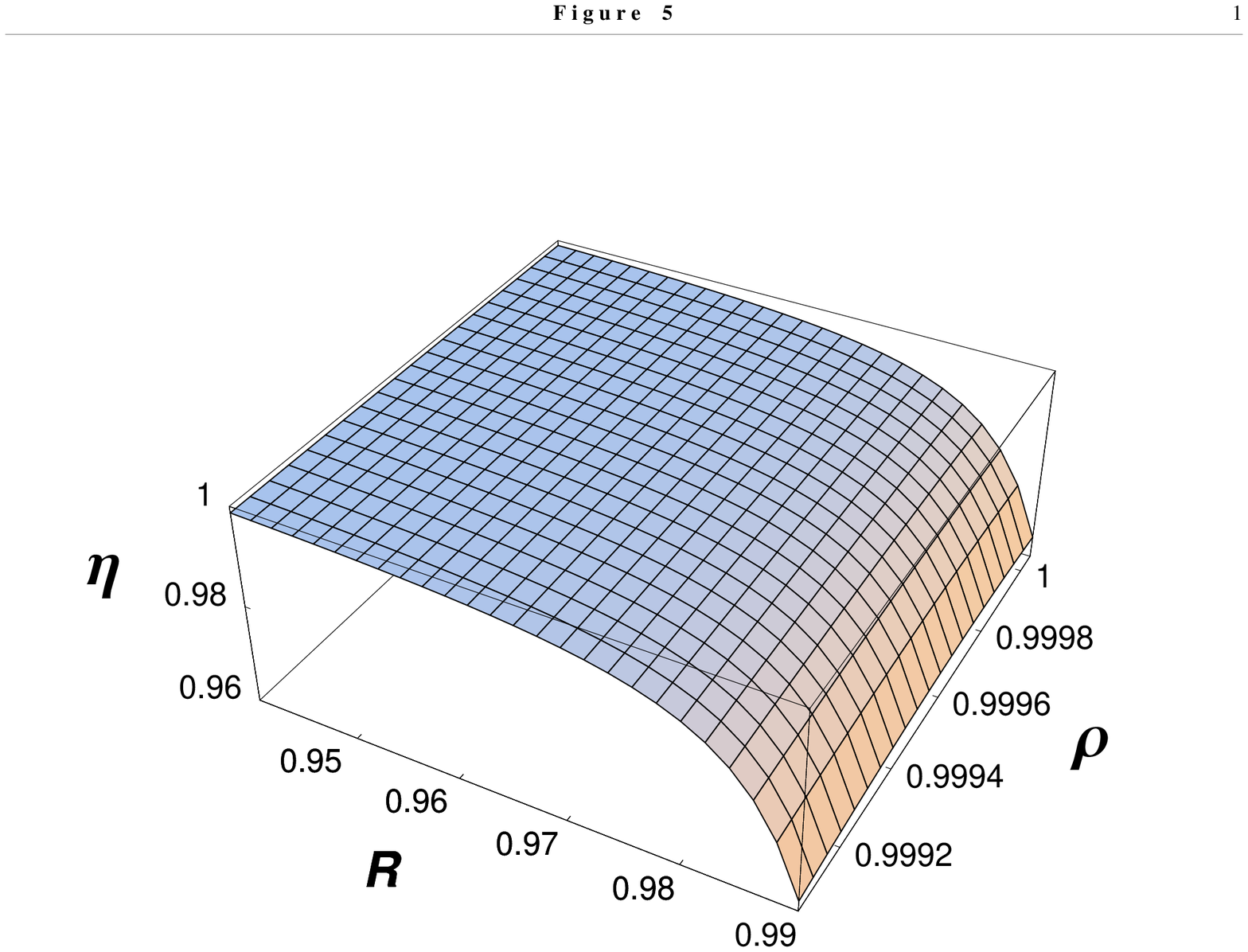}

\end{figure}

\vfill

\eject 

\vfill

\begin{figure}

\includegraphics{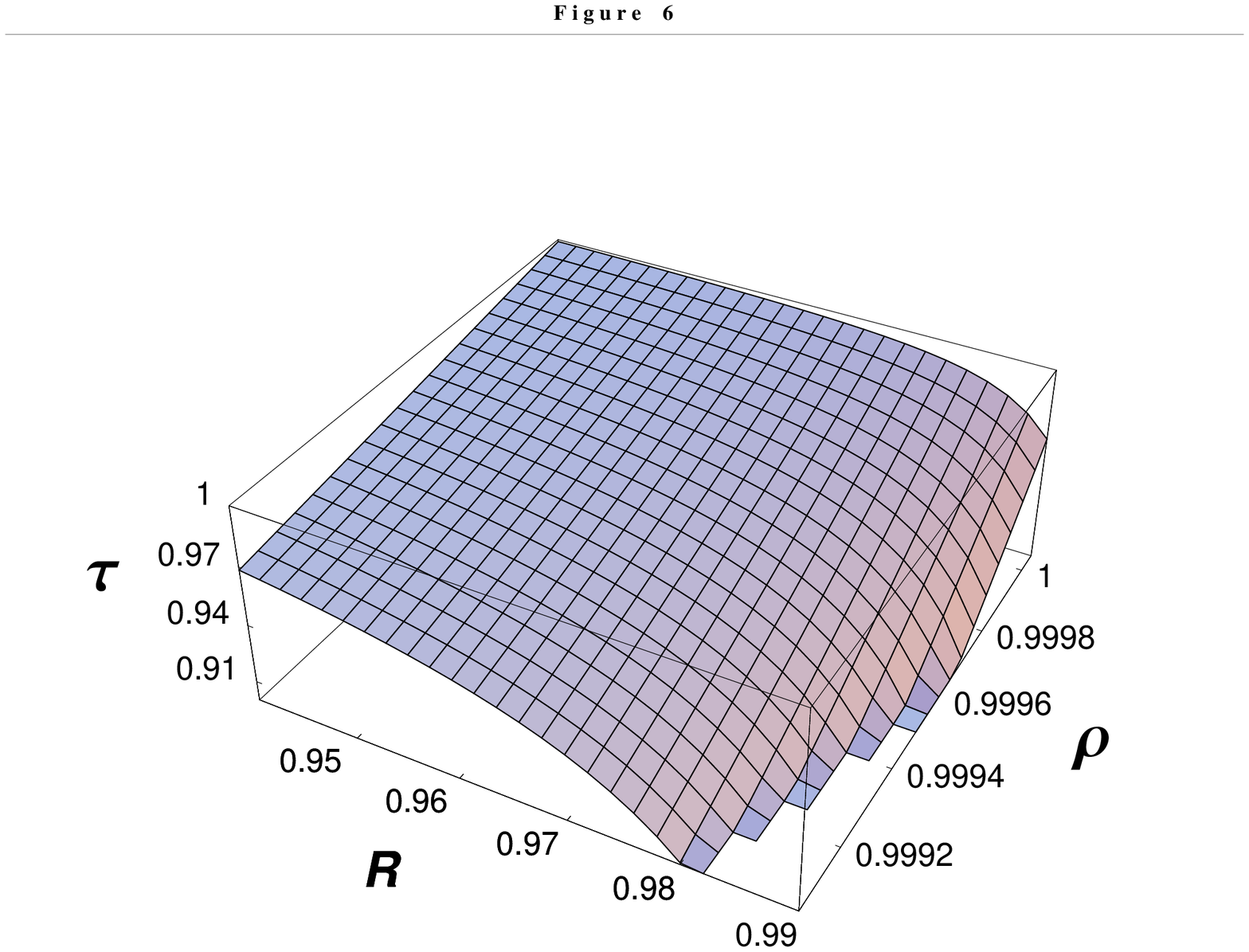}

\end{figure}

\vfill

\end{document}